\documentclass[%
 reprint,
 amsmath,amssymb,
 aps,
]{revtex4-2}

\usepackage{ytableau}
\usepackage[hyperfootnotes=false]{hyperref}
\hypersetup{
 colorlinks=true,
 citecolor=blue,
 linkcolor=red,
 urlcolor=blue}
\usepackage{mathtools}
\usepackage{nccmath} 
\usepackage{xcolor}
\usepackage{caption}
\usepackage{subcaption}
\usepackage{graphicx}
\usepackage{dcolumn,physics}
\usepackage{tabularx}
\usepackage{bm}

\begin{document}

\preprint{APS/123-QED}

\title{Spectroscopic Analysis of Singly Heavy Pentaquarks in the Symmetric 15-Plet Representation Using Phenomenological Models.}

\author{Ankush Sharma, Alka Upadhyay}
 \affiliation{Department of Physics and Material Science, Thapar Institute of Engineering and Technology, Patiala, India, 147004}
 
\email{ankushsharma2540.as@gmail.com}

\date{\today}

\begin{abstract}
We analyze the ground state pentaquark structures with a single heavy quark ($qqqq\Bar{Q}$) using various phenomenological models. The recent observations of singly heavy tetraquark structures at LHCb serve as a significant motivation for this investigation. We studied the symmetric 15-plet configuration of SU(3) flavor representation with the spin-parity assignment of $5/2^-$, representing the symmetric spin state for the pentaquark systems. We employed an extended Gursey-Radicati mass formula and an effective mass scheme to compute the mass spectra of pentaquark states. Additionally, the methodology of the screened charge scheme is introduced to calculate the magnetic moment assignments, specifically for configurations involving both charm and bottom quarks. We also proposed the potential production modes originating from the weak decay of heavy baryons. We identified the strong decay channels where pentaquark transitions into a light baryon and a heavy meson. Our analysis of mass spectra, magnetic moments, and possible strong decay channels helps us to explore the inner structure of pentaquarks and their underlying quark dynamics. This work not only augments the theoretical frameworks used to describe such systems but is also helpful for future experimental pursuits at facilities like LHCb, fostering further experimental validations and discoveries in heavy quark spectroscopy.
\end{abstract}

\keywords{Singly heavy pentaquarks, Special unitary representation, Masses, Magnetic Moments.}

\maketitle

\section{Introduction}
The study of exotic hadrons has seen remarkable progress in recent years, highlighted by the discovery of several tetraquark and pentaquark structures. The term pentaquark was first proposed in 1987 when it was found that pentaquarks with a single charm quark with quark contents $uuds\bar{c}$ and $udds\bar{c}$ and their bottom analogs are likely to be stable states \cite{GIGNOUX, LIPKIN}. These findings, achieved through high-precision experiments at facilities such as the Large Hadron Collider (LHC), have significantly broadened our understanding of Quantum Chromodynamics (QCD) in the non-perturbative regime. In 2022, the LHCb experiment identified two singly heavy tetraquarks with quark content $u\bar{d}c\bar{s}$ and $\bar{u}dc\bar{s}$ with the significance of 6.5 $\sigma$ and 8 $\sigma$ respectively \cite{2900}. These two states $T_{c\bar{s}0}^a (2900)^{++}$ and $T_{c\bar{s}0}^a(2900)^0$ were observed in the $D_s^+\pi^+$ and $D_s^+\pi^-$ invariant mass spectra in two $B$-decay processes $B^+ \rightarrow D^-D_s^+\pi^+$ and $B^0 \rightarrow \bar{D^0}D_s^+ \pi^-$, respectively with $(I)J^P = (1)0^+$. These two states correspond to the isospin triplet with mass and width as  $M_{exp} = 2908\pm 11 \pm 20$ MeV, and $\Gamma_{exp}= 136 \pm 23 \pm 13$  MeV, respectively. Moreover, way back in 2004, the H1 Collaboration reported the narrow resonance $\Theta_c$ in $D^{*-}p$ ($uudd\Bar{c}$) and $D^{*+}\Bar{p}$ ($\Bar{u}\Bar{u}\Bar{d}\Bar{d}c$) states produced in inelastic electron-proton collisions, with mass of $3099\pm 3 \pm 5$ MeV and widths $12 \pm 3$ MeV, respectively \cite{Experimental}. Various models also supported the argument that instead of a strange anti-quark ($\Bar{s}$) in the $\Theta^+$ state, a heavy antiquark ($\Bar{c}$ or $\Bar{b}$) is more likely to be bound \cite{Support1, Support2}. It was further supported by the different theoretical models that singly heavy pentaquark states are more likely to be found \cite{theory1,theory2}. Furthermore, in Ref. \cite{PRD}, an interesting phenomenon was discussed to observe the pentaquarks with configuration $qqqq\bar{Q}$ by generating a pair of strange quarks from an anti-strange quark of observed tetraquark state ($X(2900)$) by the LHCb collaboration. Furthermore, many hidden-charm pentaquark structures have been discovered in recent decades. Recently, in 2022, LHCb discovered the hidden-charm pentaquark structure, $P_{\psi s}^\Lambda(4338)^0$ with quark content $udsc\overline{c}$ observed in the $J/\psi\Lambda$ mass in the $B^-\rightarrow J/\psi\Lambda p$ decays with the statistical significance of 15$\sigma$  with a mass and width of $4338.2\pm0.7\pm0.4$ MeV and $7.0\pm1.2\pm1.3$ MeV, respectively \cite{4338}. In 2021, the hidden-charm strange pentaquark state $P_{cs}(4459)$ was observed in the J/$\psi\Lambda$ invariant mass distribution from an amplitude analysis of the $\Xi_b^- \rightarrow J/\psi \Lambda K^-$ decays, characterized by a mass of $4458.8\pm2.9_{-1.1}^{+4.7}$ MeV and a width of $17.3\pm6.5_{-5.7}^{+8.0}$ MeV \cite{20211278}. in 2015, the LHCb collaboration reported two hidden-charm structures, $P_c(4380)^+$, and $P_c(4450)^+$, in $\Lambda_b$ decay with a mass $4380\pm8\pm28$ MeV and $4449.8\pm1.7\pm2.5$ MeV with corresponding widths of $205\pm18\pm86$ MeV and $39\pm5\pm19$ MeV respectively \cite{PhysRevLett.115.072001}. These observations paved the way for exploring the analogous structures in pentaquark systems. Therefore, by taking motivation from these observations, we investigate the symmetric 15-plet configuration of SU(3) flavor representation in the context of singly charm and bottom pentaquarks. The symmetric 15-plet representation offers a rich structure that can accommodate various pentaquark configurations, including those with a single heavy quark, making it an ideal framework for our exploration. The discovery of pentaquark states has challenged the traditional quark model, necessitating a thorough investigation of their properties. By focusing on the 15-plet configuration, we aim to elucidate the role of SU(3) flavor symmetry in the formation and stability of these states. These pentaquark states, which extend beyond the conventional three-quark baryons and quark-antiquark mesons, present a unique laboratory for probing the interplay between quark confinement and chiral symmetry breaking in Quantum Chromodynamics (QCD). We employed phenomenological models like the Gursey-Radicati mass formula, effective mass scheme, and screened charge schemes to examine the mass spectra and magnetic moments of pentaquark states. Theoretical predictions of the masses and magnetic moments of pentaquarks are essential for guiding future lattice QCD simulations and experimental searches. Our approach integrates insights from heavy quark effective theory (HQET) and chiral perturbation theory (ChPT) to account for the dynamics of heavy and light quarks within these multi-quark systems. By investigating the symmetric 15-plet representation, we aim to uncover new pentaquark states with a single heavy quark that could be observed in future experiments. Additionally, understanding the properties and interactions of these pentaquarks contributes to the broader effort of mapping out the spectrum of hadronic matter predicted by QCD. Our findings provide predictions that can be tested experimentally, guiding future searches and enhancing the interpretative framework for observed pentaquark candidates. \\
The organization of this work is as follows: section II describes the theoretical framework for pentaquarks with single heavy quarks, including the classification scheme using the special unitary representations and the Young tableau technique, an extension of the Gursey-Radicati mass formula, the effective mass scheme, and the screened charge scheme to calculate the mass spectra and magnetic moments of pentaquarks. Section III consists of the analysis part for both of the 15-plet configurations for pentaquarks containing charm and bottom quarks, respectively, and section IV summarise the article.

\section{Theoretical Formalism}
Classifying pentaquark states, particularly those with a single heavy quark, is a critical step in understanding their internal structure and predicting their properties. Using the special unitary representation, we classified the pentaquarks with single heavy quarks. SU(3) flavor symmetry arises from the near degeneracy of the light quarks and is a powerful tool in classifying hadrons. For a system like the pentaquark, which consists of four quarks and one antiquark, the possible combinations of quark flavors can be organized into multiplets corresponding to specific SU(3) representations. In the case of singly heavy pentaquarks, where one of the quarks is a heavy quark (such as charm or bottom), the SU(3) symmetry applies only to the light quarks. The heavy quark acts as a spectator, and the classification focuses on the remaining four light quarks and the antiquark. Therefore, by using the SU(3) flavor representation, which assigns each light quark and light antiquark by the fundamental [3] and [$\Bar{3}$] representation, respectively, the classification scheme for $qqqq\Bar{Q}$ pentaquarks are:
\begin{align}
[3]\otimes [3]\otimes [3]\otimes [3] =  3 [3] \oplus 2 [\Bar{6}] \oplus 3 [15^{'}] \oplus[15]
\label{1}
\end{align}
Similarly, in SU(2) spin representation, each quark and anti-quark is assigned by fundamental [2] representation. Therefore, classification scheme for spin wavefunction of $qqqq\Bar{Q}$ -type pentaquarks are:
 \begin{align}
    [2] \otimes [2] \otimes [2] \otimes [2] \otimes
[2] =  [6]  \oplus 4[4] \oplus 5[2]
\label{2}
\end{align}
Here, configuration [6] corresponds to spin-5/2, [4] corresponds to spin-3/2, and [2] corresponds to spin-1/2 pentaquarks. By using Eq. \ref{1} and \ref{2}, we classified the $qqqq\Bar{Q}$- type pentaquarks into the symmetric 15-plet representation with symmetric spin-parity $J^P = 5/2^-$ representation. Young Tableau representation for symmetric 15-plet flavor representation is represented as:
\begin{center}
\begin{ytableau}
    \none &  &  &  &  
\end{ytableau}
\end{center}
This 15-plet representation is symmetric in nature. Understanding the symmetric 15-plet pentaquarks holds significant implications for experimental and theoretical high-energy physics.  Theoretically, studying these exotic states can shed light on the dynamics of quark confinement and the role of flavor symmetries in hadron spectroscopy.
The corresponding symmetric flavor wavefunction can be defined for $uuuu\Bar{c}$ as:
\begin{equation}
    \psi_{flavor} = \frac{1}{\sqrt{24}}\sum_p P(u_1u_2u_3u_4)\Bar{c}
\end{equation}
$P$ denotes all permutations of the four up quarks, and $p$ represents the permutation operator. Since all the quarks are the same (all up quarks), the wavefunction simplifies significantly due to the identical nature of the quarks. Thus, for $uuuu\Bar{c}$ pentaquark, the symmetrized flavor wavefunction becomes:
\begin{widetext}
\begin{align}
    \psi_{flavor} =& \frac{1}{\sqrt{24}}(u_1u_2u_3u_4 + u_1u_2u_4u_3 + u_1u_3u_2u_4 + u_1u_3u_4u_2 + u_1u_4u_2u_3 + u_1u_4u_3u_2 + u_2u_1u_3u_4 + u_2u_1u_4u_3 \\ \nonumber &+ u_2u_3u_1u_4 + u_2u_3u_4u_1 + u_2u_4u_1u_3 + u_2u_4u_3u_1 + u_3u_1u_2u_4 + u_3u_1u_4u_2 + u_3u_2u_1u_4 + u_3u_2u_4u_1 \\ \nonumber &+ u_3u_4u_1u
_2 + u_3u_4u_2u_1 + u_4u_1u_2u_3 + u_4u_1u_3u_2 + u_4u_2u_1u_3 + u_4u_2u_3u_1 + u_4u_3u_1u_2 + u_4u_3u_2u_1) \Bar{c}
\end{align}
 \end{widetext} 
This symmetrization ensures that the flavor wavefunction is symmetric under the exchange of any two up quarks. Similarly, we can define this for other members of the 15-plet representation. Moreover, for spin-5/2 pentaquarks, Young tableau representations are:
 \begin{center}
  \begin{ytableau}
    \none & 1 & 2 & 3 & 4 & 5
\end{ytableau}
\end{center}
and the corresponding spin wavefunction for spin $5/2^-$ is $\ket{\uparrow\uparrow\uparrow\uparrow\uparrow}$. The wavefunction $\ket{\uparrow\uparrow\uparrow\uparrow\uparrow}$ describes a completely symmetric spin state under permutations of the quarks. It is fully symmetric because exchanging any two quarks leaves the wavefunction unchanged. This is a requirement for the wavefunction to be consistent with the Pauli exclusion principle, which governs the overall symmetry properties of fermions. However, because the wavefunction represents a system with quarks (which are fermions), the overall wavefunction of the pentaquark must be antisymmetric under the exchange of any two identical quarks. As we study the ground state pentaquark states, the symmetrical restriction for the spatial wave function is trivial. The color wave function associated with a bound state of a hadron is subject to color confinement, resulting in color singlet states. Hence, the overall wavefunction of the 15-plet pentaquark states is antisymmetric with symmetric spin, flavor, and spatial parts of the wavefunction, and the resulting anti-symmetry arises from the color wavefunction. In the next subsection, the extended version of the Gursey-Radicati mass formula is introduced to calculate the mass spectra of pentaquark states.

\subsection{The Gursey-Radicati mass formula}
The Gursey-Radicati mass formula was initially proposed to describe the mass splittings within multiplets of baryons and mesons, incorporating contributions from various quantum numbers, such as spin, isospin, and flavor \cite{FG}. In its extended form, this formula can be adapted to account for the more complex structure of pentaquarks. The formalism of the Gursey-Radicati mass formula has proven to be an effective method for calculating the masses of exotic hadrons in recent times \cite{Santo, HOLMA, Sharma_2023, Ankush, sharma2024}. Firstly, it was introduced to study the mass spectra of baryons. Later, it was modified to study the mass spectra of exotic hadrons by introducing the counter term and modifying the scale parameter as \cite{Santo}:
\begin{align}
  M_{GR} = M_0 &+ AS(S+1) + DY  + E[I(I+1) -1/4 Y^2] \nonumber \\ &+ G C_2(SU(3)) + F N_c
\end{align}
This formula was further modified by including the contribution from the bottom quark by generalizing the counter term \cite{HOLMA}:
 \begin{align}
  M_{GR} = \xi M_0 &+ AS(S+1) + DY  + E[I(I+1)  -1/4 Y^2] \nonumber \\ &+ G C_2(SU(3)) + \sum_{i=c,b} F_i N_i
  \label{mass formula}
  \end{align}
here, $M_0$ is the scale parameter, and $\xi$ is the correction factor to the scale parameter, related to the number of quarks making up the hadron. The quantum numbers spin, isospin, and hypercharge are denoted as $S$, $I$, and $Y$, respectively. $N_i$ represents the counter term for the number of heavy quarks (anti-quarks) making up the hadron. Mass formula parameters are taken from Ref. \cite{Santo, HOLMA}. Mass formula parameters are listed in Table \ref{tab:2}.

\begin{table}[ht]
\centering
\caption{Values of parameters used in the extended GR mass formula with corresponding uncertainties. \cite{Santo}}
\tabcolsep 0.4mm  
\begin{tabular}{cccccccc}
       \hline
       \hline
         & $M_0$ & A & D & E & G & $F_c$ & $F_b$ \\
         \hline
        Values[MeV] & 940.0 & 23.0 & -158.3 & 32.0 & 52.5 & 1354.6  & 4820 \cite{HOLMA} \\
        \hline
        Uncert.[MeV] & 1.5 & 1.2 & 1.3 & 1.3 & 1.3 & 18.2  & 34.4  \\
        \hline
        \hline
       \end{tabular}
        \label{tab:2}
   \end{table}
Using eq.\eqref{mass formula}, the mass spectrum of pentaquarks with single heavy quarks is calculated with respective uncertainties and reported in Tables II and III, respectively. The results presented in Tables II and III highlight several key aspects of the mass spectrum of pentaquarks with a singly charm and bottom quarks. The uncertainties associated with our calculations reflect the inherent challenges in modeling multi-quark systems and underscore the need for further experimental and theoretical studies. The observed masses suggest a strong dependence on the interplay between the light and heavy quarks, as well as the contributions from spin, isospin, hypercharge and strangeness. By incorporating the distinctive contributions of the heavy quark and the modified interactions among quarks, we achieve a more accurate and comprehensive description of pentaquark masses. In the next subsection, we defined the effective mass scheme to calculate the effective masses and magnetic moments of pentaquarks with a single heavy quark.   

\subsection{Effective mass scheme}
The effective mass scheme is a theoretical framework used to describe the masses of composite particles, such as singly heavy pentaquarks. The effective mass scheme involves modeling the mass of the pentaquark as the sum of the constituent quark masses plus contributions from the interactions between them. This approach simplifies the complex dynamics of the strong interaction within the hadron by approximating certain effects. The masses of the individual quarks, including the heavy quark (c for charm or b for bottom), are considered in the effective mass. These effective masses are usually larger than the bare quark masses due to the inclusion of effects such as the quark’s interaction with the QCD vacuum. The effective mass scheme includes terms that account for the interaction energies between quarks. These interactions are typically modeled using potential terms derived from QCD-inspired models, such as one-gluon exchange or chiral interactions. The spin-spin interactions between quarks, particularly between the heavy quark and the light quarks, can contribute significantly to the mass. These interactions are often modeled using a hyperfine interaction term proportional to the product of the spins of the quarks. Thus, we can define the mass of the pentaquark as \cite{Khanna, VERMA}:
\begin{align}
        M_P =& \sum_{i=1}^5 m_i^{eff}  \\
   M_P =& \sum_{i=1}^5 m_i + \sum_{i<j} b_{ij} s_i.s_j
   \label{Effective mass}
\end{align}
 Here, $s_i$ and $s_j$ represent spin operators for the $i^{th}$ and $j^{th}$ quarks (antiquarks), and $m_i^{eff}$ are the effective mass for each quark (antiquarks) and $b_{ij}$ are the hyperfine interaction parameters. The effective masses equations for constituent quarks making up the pentaquarks are:
 \begin{equation}
 m_1^{eff} = m_1 + \alpha b_{12} + \beta b_{13} + \gamma b_{14} + \eta b_{15}
 \end{equation}
 
 \begin{equation}
  m_2^{eff} = m_2 + \alpha b_{12} + \beta^{'} b_{23} + \gamma^{'} b_{24} + \eta^{'} b_{25}
  \end{equation}

   \begin{equation}
   m_3^{eff} = m_3 + \beta b_{13} + \beta^{'} b_{23} + \gamma^{''} b_{34} + \eta^{''} b_{35} 
   \end{equation}

   \begin{equation}
    m_4^{eff} = m_4 + \gamma b_{14} + \gamma^{'} b_{24} + \gamma^{''} b_{34} + \eta^{'''} b_{45} 
 \end{equation}

  \begin{equation}
    m_5^{eff} = m_5 + \eta b_{15} + \eta^{'} b_{24} + \eta^{''} b_{34} + \eta^{'''} b_{45} 
 \end{equation}
  Here, numbers 1, 2, 3, 4, and 5 stand for quarks such as $u$, $d$, $s$, $c$, and $b$. These effective mass equations further get modified if we include more number of identical quarks \cite{Rohit}. 
  \begin{equation}
s_1.s_2 = s_2.s_3 =  s_3.s_4 =
 s_4.s_5 = 1/4 
\end{equation}
 and parameters are calculated by using \eqref{Effective mass}:
 \begin{equation}
     \alpha = \beta = \gamma = \eta = 1/8
 \end{equation}

\begin{equation}
\beta^{'} = \gamma^{'} = \eta^{'} = 1/8
\end{equation}

\begin{equation}
\gamma^{''} = \eta^{''} = \eta^{'''} = 1/8
\end{equation}
Therefore, the effective mass equation for $J^P = 5/2^-$ ($\uparrow\uparrow\uparrow\uparrow\uparrow$) defined as:
\begin{align}
M_{P_{{5/2}^-}} = m_1 + m_2 + m_3 + m_4 + m_5 + \frac{b_{12}}{4}  + \frac{b_{13}}{4}  + \frac{b_{14}}{4} \nonumber \\
 + \frac{b_{15}}{4}  + \frac{b_{23}}{4}  + \frac{b_{24}}{4}  + \frac{b_{25}}{4}  + \frac{b_{34}}{4}  + \frac{b_{35}}{4}  + \frac{b_{45}}{4}
 \label{eff1}
\end{align}

Masses of respective quarks are taken from Ref. \cite{Rohit}. Hyperfine interaction terms are calculated using the effective mass equation \eqref{eff1} by taking the masses obtained from the Gursey-Radicati mass formula as a reference:
\begin{align}
     m_u = m_d &= \hspace{0.3cm} 362 MeV, \hspace{0.3cm} m_s &= \hspace{0.3cm} 539 MeV \nonumber \\
    m_c &= \hspace{0.3cm} 1710 MeV, \hspace{0.3cm} m_b &= \hspace{0.3cm} 5043 MeV
\end{align}
\begin{equation}
     b_{uu} = \hspace{0.3cm} b_{ud} = \hspace{0.3cm} b_{dd} = 157.74 MeV
 \end{equation}

 \begin{equation}
     b_{us} = \hspace{0.3cm} b_{ds} \hspace{0.3cm} = \left(\frac{m_u}{m_s}\right)b_{uu} = 105.94 MeV
 \end{equation}

\begin{equation}
     b_{uc} = \hspace{0.3cm} b_{dc} = \hspace{0.3cm} \left(\frac{m_u}{m_c}\right)b_{uu} = 33.21 MeV
 \end{equation}

\begin{equation}
     b_{ss} = \hspace{0.3cm} \left(\frac{m_u}{m_s}\right)^2 b_{uu} = 71.50  MeV
 \end{equation}

\begin{equation}
     b_{sc} = \hspace{0.3cm} \left(\frac{m_u^2}{m_s m_c}\right) b_{uu} = 22.42 MeV
\end{equation}

\begin{equation}
     b_{ub} = \hspace{0.3cm} b_{db} = \hspace{0.3cm} \left(\frac{m_u}{m_b}\right)b_{uu}  = 11.33 MeV
 \end{equation}

 \begin{equation}
     b_{sb} = \hspace{0.3cm} \left(\frac{m_u}{m_s}\right) b_{ub} =  7.60 MeV
 \end{equation}
 By incorporating these hyperfine interaction terms into the total effective mass equation \eqref{eff1}, we initially determined the effective masses of the constituent quarks. Subsequently, these effective quark masses were substituted into the total effective mass equations, allowing us to compute the corresponding masses and magnetic moments for each configuration. The effective quark masses for configurations of singly charm and bottom configurations with several strange quarks are shown as follows:\\
 \\
\textbf{For $qqqq\bar{c}$ 15-plet Pentaquarks}\\

i) For ($s$ = 0) pentaquarks,
\begin{equation*}
    m_u^* = \hspace{0.3cm} m_d^* = 425.30 MeV
    \end{equation*}
    \begin{equation*}
     m_{\Bar{c}}^* = 1726.60 MeV
    \end{equation*}
\\
(ii) For ($s$ = 1) pentaquarks,
\begin{equation*}
    m_u^* = \hspace{0.3cm} m_d^* = 418.82 MeV
    \end{equation*}
    \begin{equation*}
        m_s^* = \hspace{0.3cm} 581.53 MeV
    \end{equation*}
    \begin{equation*}
     m_{\Bar{c}}^* = 1725.56 MeV
    \end{equation*}
\\

(iii) For ($s$ = 2) pentaquarks,
\begin{equation*}
    m_u^* = \hspace{0.3cm} m_d^* = 412.00 MeV
    \end{equation*}
    \begin{equation*}
        m_s^* = \hspace{0.3cm} 577.00 MeV
    \end{equation*}
    \begin{equation*}
    m_{\Bar{c}}^* = 1723.00 MeV
    \end{equation*}
\\

(iv) For ($s = 3$) pentaquarks,
\begin{equation*}
    m_u^* = \hspace{0.3cm} m_d^* = 405.87 MeV
    \end{equation*}
    \begin{equation*}
        m_s^* = \hspace{0.3cm} 572.92 MeV
    \end{equation*}
    \begin{equation*}
     m_{\Bar{c}}^* = 1722.56 MeV
    \end{equation*}
\\
(v) For ($s = 4$) pentaquarks,
\begin{equation*}
        m_s^* = \hspace{0.3cm} 577.00 MeV
    \end{equation*}
    \begin{equation*}
     m_{\Bar{c}}^* = 1721.00 MeV
    \end{equation*}
\\
\textbf{For $qqqq\bar{b}$ 15-plet Pentaquarks}\\

i) For ($s$ = 0) pentaquarks,
\begin{equation*}
    m_u^* = \hspace{0.3cm} m_d^* = 422.00 MeV
    \end{equation*}
    \begin{equation*}
     m_{\Bar{b}}^* = 5048.00 MeV
    \end{equation*}
\\
(ii) For ($s$ = 1) pentaquarks,
\begin{equation*}
    m_u^* = \hspace{0.3cm} m_d^* = 416.00 MeV
    \end{equation*}
    \begin{equation*}
        m_s^* = \hspace{0.3cm} 579.00 MeV
    \end{equation*}
    \begin{equation*}
     m_{\Bar{b}}^* = 5048.00 MeV
    \end{equation*}
\\

(iii) For ($s$ = 2) pentaquarks,
\begin{equation*}
    m_u^* = \hspace{0.3cm} m_d^* = 409.61 MeV
    \end{equation*}
    \begin{equation*}
        m_s^* = \hspace{0.3cm} 575.37 MeV
    \end{equation*}
    \begin{equation*}
    m_{\Bar{b}}^* = 5047.00 MeV
    \end{equation*}
\\

(iv) For ($s = 3$) pentaquarks,
\begin{equation*}
    m_u^* = \hspace{0.3cm} m_d^* = 403.00 MeV
    \end{equation*}
    \begin{equation*}
        m_s^* = \hspace{0.3cm} 571.00MeV
    \end{equation*}
    \begin{equation*}
     m_{\Bar{b}}^* = 5047.00 MeV
    \end{equation*}
\\
(v) For ($s = 4$) pentaquarks,
\begin{equation*}
        m_s^* = \hspace{0.3cm} 566.00 MeV
    \end{equation*}
    \begin{equation*}
     m_{\Bar{b}}^* = 5046.00 MeV
    \end{equation*}

\subsection{Screened charge scheme}
The concept of a screened charge scheme in the context of singly heavy pentaquarks relates to how the internal quark structure and interactions are described, particularly concerning the screening effects that influence the effective charge and potential between quarks. The term "screened charge" refers to the phenomenon where the effective charge felt by quarks within a hadron is modified due to the presence of other quarks and the gluonic field. The screened charge scheme is crucial for understanding the internal dynamics and properties of singly heavy pentaquarks. It provides a framework to describe how the quarks within these exotic states interact with each other, influencing their stability and the spectrum of possible pentaquark states. By incorporating screening effects, theoretical models and simulations can more accurately predict the behavior of these complex hadronic systems, guiding experimental searches and interpretations of pentaquark observations. By incorporating screening effects, theoretical models and simulations can more accurately predict the behavior of these complex hadronic systems, guiding experimental searches and interpretations of pentaquark observations. The screening effect plays a significant role in determining the binding energy and stability of pentaquarks. A proper understanding of the screened charge is essential to predict whether a given pentaquark state is stable or if it will decay quickly.
The effective reduction in the interaction strength due to screening can lead to more tightly bound states or, conversely, to states that are more loosely bound and thus more prone to decay. the effective charge of the quark '$q$' in pentaquark ($q$, $r$, $v$, $x$, $y$) can be written as: 
\begin{equation}
    e_q^P = e_q + \alpha_{qr} e_r + \alpha_{qv} e_v + \alpha_{qx} e_x + \alpha_{qy} e_y
    \label{screen1}
    \end{equation}
 Here $e_q$ stands for the charge of the quark (anti-quark). Similarly, we can define these equations for other quarks (anti-quarks) inside the pentaquarks \cite{sharma2024}. By considering the isospin symmetry:
\begin{align}
    \alpha_{uu} = \alpha_{ud} = \alpha_{dd} = \alpha_1 \\ \nonumber
    \alpha_{us} = \alpha_{ds} = \beta_1 \\ \nonumber
    \alpha_{ss} = \beta_2
\end{align}
 for states with charm quark,
 \begin{align}
     \alpha_{uc} = \alpha_{dc} = \beta_3 \\ \nonumber
     \alpha_{sc} = \alpha_2, \hspace{0.3cm} \alpha_{cc} = \alpha_3
 \end{align}
Similarly, for the states with bottom quark,
 \begin{align}
     \alpha_{ub} = \alpha_{db} = \beta_4, \hspace{0.3cm} \alpha_{sb} = \alpha_4 \\ \nonumber
     \alpha_{cb} = \beta_5, \hspace{0.3cm} \alpha_{bb} = \alpha_5
 \end{align}
by using the SU(3) symmetry, these parameters can be reduced as:
\begin{equation}
    \alpha_1 = \beta_1 = \beta_2
\end{equation}
For the calculation of screening parameter $\alpha_{ij}$, we used the Ansatz formalism, 
\begin{equation}
    \alpha_{ij} = \mid{\frac{m_i - m_j}{m_i + m_j}}\mid \times \delta
\end{equation}
$m_i$ and $m_j$ stands for the respective quark masses and $\delta$ = 0.81 \cite{Bains}. The screening parameters help us predict the magnetic moments of pentaquarks. By substituting these parameters in effective charge equations, we can calculate the magnetic moments of pentaquarks by introducing the magnetic moment operator:
\begin{equation}
    \mu = \sum_i \frac{e_i^P}{2 m_i^{eff}} \sigma_i
\end{equation}
The magnetic moment operator consists of two parts:
 \begin{equation}
     \Vec{\mu} = \Vec{\mu}_{spin} + \Vec{\mu}_{orbit}
 \end{equation}
which is defined as:
\begin{equation}
    \Vec{\mu} = \hspace{0.3cm} \sum_i \mu_i (2 \Vec{s_i} + \Vec{l_i}) = \hspace{0.3cm} \sum_i \mu_i(2\Vec{s_i}) = \hspace{0.3cm} \sum_i \mu_i(\Vec{\sigma_i})
\label{Magnetic moment}
\end{equation}
Since we are studying the ground states of pentaquarks with a single heavy quark, the magnetic moment depends only on the spin part. Therefore, by calculating the expectation value of Eq.\eqref{Magnetic moment} using the spin-flavor wavefunction of pentaquark with a single heavy quark, magnetic moments can be obtained as:
\begin{equation}
    \mu = \bra{\Psi_{sf}}\Vec{\mu}\ket{\Psi_{sf}}
    \label{magnetic123}
\end{equation}
Using the group theory approach, the spin-flavor wavefunction of the pentaquark is constructed and can be written as:
 \begin{equation}
     \Psi_{sf} =  \chi_s \otimes \phi_f 
 \end{equation}
The spin wavefunction for pentaquarks with $J^P = 5/2^-$ is defined as:
\begin{center}
    $\chi_s = \ket{\uparrow\uparrow\uparrow\uparrow\uparrow}$
\end{center}
and flavor wavefunction for one of the states of the 15-plet, i.e. $\ket{uuuu\Bar{c}}$ is defined as:
\begin{widetext}
\begin{align}
    \phi_{f} =& \frac{1}{\sqrt{24}}(u_1u_2u_3u_4 + u_1u_2u_4u_3 + u_1u_3u_2u_4 + u_1u_3u_4u_2 + u_1u_4u_2u_3 + u_1u_4u_3u_2 + u_2u_1u_3u_4 + u_2u_1u_4u_3 \\ \nonumber &+ u_2u_3u_1u_4 + u_2u_3u_4u_1 + u_2u_4u_1u_3 + u_2u_4u_3u_1 + u_3u_1u_2u_4 + u_3u_1u_4u_2 + u_3u_2u_1u_4 + u_3u_2u_4u_1 \\ \nonumber &+ u_3u_4u_1u
_2 + u_3u_4u_2u_1 + u_4u_1u_2u_3 + u_4u_1u_3u_2 + u_4u_2u_1u_3 + u_4u_2u_3u_1 + u_4u_3u_1u_2 + u_4u_3u_2u_1) \Bar{c}
\end{align}
 \end{widetext} 
Similarly, we can define this for other 15-plet particles of both charm and bottom states. Using the spin-flavor wavefunction in equation \eqref{magnetic123}, magnetic moments of the 15-plet states of pentaquarks with single heavy quarks are estimated. The expressions for magnetic moments are written in Table \ref{tab: expressions}, and the results are written in Table \ref{tab: magnetic moments}.

\section{Analysis of the Symmetric 15-Plet Pentaquark States with a single heavy quark.}
In exploring exotic hadronic states, pentaquarks got significant attention due to their unique quark compositions and potential to unveil new aspects of Quantum Chromodynamics (QCD). Specifically, the symmetric 15-plet pentaquark states containing a single heavy quark present an interesting subject for theoretical and experimental investigations. These states, characterized by their quark configuration and symmetry properties, offer a rich landscape for studying the interplay between the heavy quark and the light quark system. The analysis of these pentaquark states involves evaluating their mass spectra and magnetic moment assignments. By using theoretical tools such as the extended Gursey-Radicati mass formula, effective mass schemes, and screened charge models, we aim to predict the physical properties of these pentaquarks. The results obtained from this analysis not only enhance our understanding of the symmetric 15-plet states but also contribute to the broader effort of mapping out the spectrum of exotic hadrons in the context of Quantum Chromodynamics (QCD). The detailed analysis of the singly charm and bottom pentaquark states are as follows.

\subsection{$qqqq\Bar{c}$ pentaquarks}
\begin{figure}[ht!]
\centering
\caption{$Y-I_3$ plot for the 15-plet representation of $qqqq\bar{c}$ pentaquark states having single charm quark. Each state is represented by the corresponding quark content.}
\includegraphics[width=1.1\linewidth]{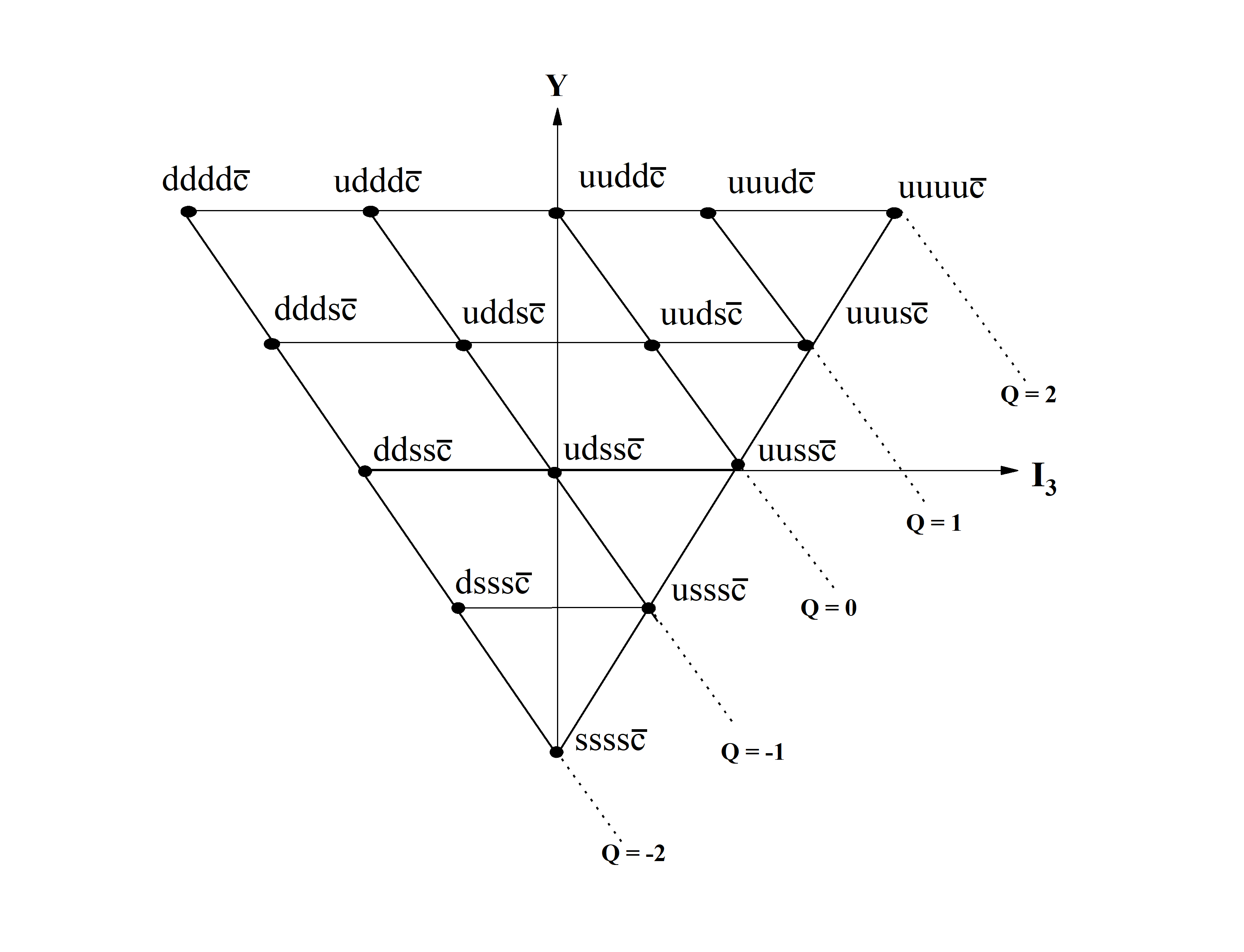}
\label{fig:1}
\end{figure}

In this subsection, we focus on the classification and properties of singly charm pentaquarks within the symmetric 15-plet representation of SU(3) flavor symmetry. The pentaquarks with the quark content ($qqqq\bar{c}$), where $q$ denotes a light quark ($u$, $d$, or $s$) and $\bar{c}$ represent a charm antiquark, are systematically categorized into the symmetric 15-plet representation. This representation is characterized by the SU(3) flavor labels (p, q) = (4, 0), as depicted in Figure \ref{fig:1}. Utilizing the extension of the Gursey-Radicati mass formula appropriately, we derive the masses for the 15 states within this representation. The effective mass scheme accounts for the various mass contributions from the constituent quarks and their interactions. The calculated masses of these pentaquark states are summarized in Table \ref{tab:II}, providing a detailed prediction of their spectral properties. Table \ref{tab:II} presents the calculated masses of the symmetric 15-plet pentaquark states containing a single charm quark. The masses are computed using the extended Gursey-Radicati mass formula and the effective mass scheme, which are essential theoretical tools in predicting the properties of exotic hadrons. The GR mass formula provides a set of predicted mass values for each pentaquark state, with associated uncertainties, highlighting the theoretical expectations for these states within a framework that extends traditional quark models. The effective mass scheme, on the other hand, offers an alternative approach that simplifies the calculations by incorporating effective quark masses, leading to a more direct estimation of the total pentaquark mass.
Additionally, we determine the magnetic moments of these states using the screened charge scheme. The corresponding expression for the magnetic moments and values are shown in  Tables \ref{tab: expressions}  and \ref{tab: magnetic moments}, respectively. Table \ref{tab: magnetic moments} presents the magnetic moments of pentaquarks with a single charm quark, calculated using three distinct theoretical schemes: the effective mass scheme, the screened charge scheme, and a combination of both. This approach incorporates the effects of quark interactions and spatial distribution within the pentaquark structure, offering a more accurate estimation of magnetic moments. Analyzing the magnetic moments of singly charmed pentaquarks using three theoretical frameworks reveals important trends. It highlights the significant role of both quark content and theoretical assumptions in determining the magnetic properties of these exotic states. The effective mass scheme tends to overestimate the magnetic moments by focusing on the quark mass contributions.
In contrast, the screened charge scheme provides a more conservative estimate by accounting for charge screening effects. The combined scheme offers a balanced approach, yielding predictions that may better align with future experimental measurements. These calculations not only shed light on the intrinsic properties of the pentaquarks but also provide critical parameters for experimental identification. Our comprehensive analysis reveals the mass spectrum and magnetic moment assignments for the symmetric 15-plet of singly charm pentaquarks, contributing to the theoretical understanding and guiding future experimental searches in exotic hadrons.

\begin{table*}\renewcommand{\arraystretch}{0.8}
 \tabcolsep 0.5mm      
\centering
\caption{Table for the masses of 15-plet representation of pentaquarks with a single charm quark using the extension of the 
Gursey-Radicati mass formula and the effective mass scheme.}
\begin{tabular}{cccc}
\hline
\hline
 States & \hspace{0.3cm} G-R mass formula(MeV) & \hspace{0.3cm} Effective mass scheme & \hspace{0.3cm} Reference \\
  \hline
  \\
$P_{c0}^{++}$,  $P_{c0}^{+}$,  $P_{c0}^{0}$,  $P_{c0}^{-}$,  $P_{c0}^{--}$  & 3579.16 $\pm$ 25.41 & 3427.82 & \\
  \\
  \hline
  \\
  $P_{c1}^+$, $P_{c1}^0$, $P_{c1}^-$, $P_{c1}^{--}$  & 3678.79 $\pm$ 24.79 & 3563.27 & \\
 \\
  \hline
  \\
 $P_{c2}^{0}$,  $P_{c2}^{-}$,  $P_{c2}^{--}$ & 3778.43 $\pm$ 24.44 & 
 3703.06 & \\
  
  \\
  \hline
  \\
  $P_{c3}^{-}$, $P_{c3}^{--}$  & 3878.66 $\pm$ 24.40  & 3847.20 & \\
  \\
  \hline
  \\
  $P_{c4}^{--}$ & 3977.69 $\pm$ 24.66  & 3995.67  & \\
    \\
    \hline
    \hline
 \end{tabular}
   \label{tab:II}
\end{table*}

\begin{table}[ht]
    \centering
    \caption{Expressions for the magnetic moments of ($J^P =
5/2^-$) singly charm pentaquarks using effective quark masses (in $\mu_N$ ).}
    \begin{tabular}{cc}
    \hline
    \hline
\tabcolsep 0.5mm
 States  & \hspace{0.3cm} Magnetic moment expressions \\
 \hline
$P_{c0}^{++}$  & \hspace{0.3cm} 4 $\mu_u^{eff}$ + $\mu_{\Bar{c}}^{eff}$
\\
$P_{c0}^{+}$ & \hspace{0.3cm} 3$\mu_u^{eff}$ + $\mu_d^{eff}$ +  $\mu_{\Bar{c}}^{eff}$ \\
$P_{c0}^{0}$ & \hspace{0.3cm} 2$\mu_u^{eff}$ + 2$\mu_d^{eff}$ + $\mu_{\Bar{c}}^{eff}$ \\
$P_{c0}^{-}$ & \hspace{0.3cm} $\mu_u^{eff}$ + 3 $\mu_d^{eff}$ + $\mu_{\Bar{c}}^{eff}$  \\
$P_{c0}^{--}$ & \hspace{0.3cm} 4$\mu_d^{eff}$ + $\mu_{\Bar{c}}^{eff}$  \\
$P_{c1}^{+}$ & \hspace{0.3cm} 3$\mu_u^{eff}$ + $\mu_s^{eff}$ + $\mu_{\Bar{c}}^{eff}$ \\
$P_{c1}^{0}$ & \hspace{0.3cm} 2$\mu_u^{eff}$ + $\mu_d^{eff}$ + $\mu_s^{eff}$ + $\mu_{\Bar{c}}^{eff}$   \\
$P_{c1}^{-}$ & \hspace{0.3cm}  $\mu_u^{eff}$ + 2 $\mu_d^{eff}$ + $\mu_s^{eff}$ + $\mu_{\Bar{c}}^{eff}$  \\
$P_{c1}^{--}$ & \hspace{0.3cm} 3 $\mu_d^{eff}$ + $\mu_s^{eff}$ + $\mu_{\Bar{c}}^{eff}$  \\
$P_{c2}^{0}$ & \hspace{0.3cm}  
 2$\mu_u^{eff}$ + 2$\mu_s^{eff}$ + $\mu_{\Bar{c}}^{eff}$ \\
$P_{c2}^{-}$ & \hspace{0.3cm}  $\mu_u^{eff}$ + $\mu_d^{eff}$ + 2$\mu_s^{eff}$ + $\mu_{\Bar{c}}^{eff}$ \\
$P_{c2}^--$ & \hspace{0.3cm} 2 $\mu_d^{eff}$ + 2$\mu_s^{eff}$ + $\mu_c^{eff}$ + $\mu_{\Bar{c}}^{eff}$ \\
$P_{c3}^{-}$ & \hspace{0.3cm} $\mu_u^{eff}$ + 3$\mu_s^{eff}$ + $\mu_{\Bar{c}}^{eff}$ \\
$P_{c3}^{--}$ & \hspace{0.3cm} $\mu_d^{eff}$ +  3$\mu_s^{eff}$ + $\mu_{\Bar{c}}^{eff}$ \\
$P_{c4}^{--}$ & \hspace{0.3cm}  4$\mu_s^{eff}$ + $\mu_{\Bar{c}}^{eff}$ \\
        \hline
        \hline
    \end{tabular}
  \label{tab: expressions}
\end{table}

\begin{table*}
    \centering
    \caption{Magnetic moments of pentaquarks with single charm quark having $J^P = 5/2^-$ using the effective mass scheme, screened charge scheme, and effective mass and screened charge schemes together.}
    \begin{tabular}{|p{1cm}|c|c|c|c|}
    \hline
 States  &  Quark Content &  Effective mass Scheme  &  Screened Charge Scheme &  Effective mass + Screened Charge scheme\\
     \hline
$P_{c0}^{++}$ &  $uuuu\Bar{c}$ &  5.52 & 3.68 &  3.19\\
      
$P_{c0}^{+}$ &  $uuud\Bar{c}$ & 3.31 &  0.80 & 0.69 \\
       
$P_{c0}^{0}$ &  $uudd\Bar{c}$ &  1.10 &  -2.08 &  -1.79 \\
         
$P_{c0}^{-}$ &  $uddd\Bar{c}$ &  -1.09 & -4.96 & -4.29\\
         
$P_{c0}^{--}$ &  $dddd\Bar{c}$ &  -3.30 &  -7.84 & -6.78 \\
          
$P_{c1}^{+}$ &  $uuus\Bar{c}$ &  3.58 &  1.66 & 1.42\\
           
$P_{c1}^{0}$ &  $uuds\Bar{c}$ &  1.34 & -0.78 & -0.73 \\
            
$P_{c1}^{-}$ &  $udds\Bar{c}$ &  -0.90 & -4.65 & -4.14 \\
             
$P_{c1}^{--}$ & $ddds\Bar{c}$ & -3.14 & -7.81  & -6.92 \\
              
$P_{c2}^{0}$ & $uuss\Bar{c}$ & 1.58 & -0.44 & -0.48\\
 $P_{c2}^{-}$ &  $udss\Bar{c}$ & -0.68 & -3.88 & -3.56\\
           
 $P_{c2}^{--}$ &  $ddss\Bar{c}$ &  -2.96 &  -7.04 & -6.38 \\
            
$P_{c3}^{-}$ &  $usss\Bar{c}$ &  -0.46 & -2.64 & -2.53 \\
             
$P_{c3}^{--}$ & $dsss\Bar{c}$ & -2.77 & -1.98  & -2.01 \\
              
$P_{c4}^{--}$ & $ssss\Bar{c}$ & -2.56 & -4.94 & -4.71 \\
\hline
\end{tabular}
    \label{tab: magnetic moments}
\end{table*}

\subsection{$qqqq\Bar{b}$ pentaquarks}
\begin{figure}[ht]
\centering
\caption{$Y-I_3$ plot for the 15-plet representation of $qqqq\bar{b}$ pentaquark states having single charm quark.}
\includegraphics[width=\linewidth]{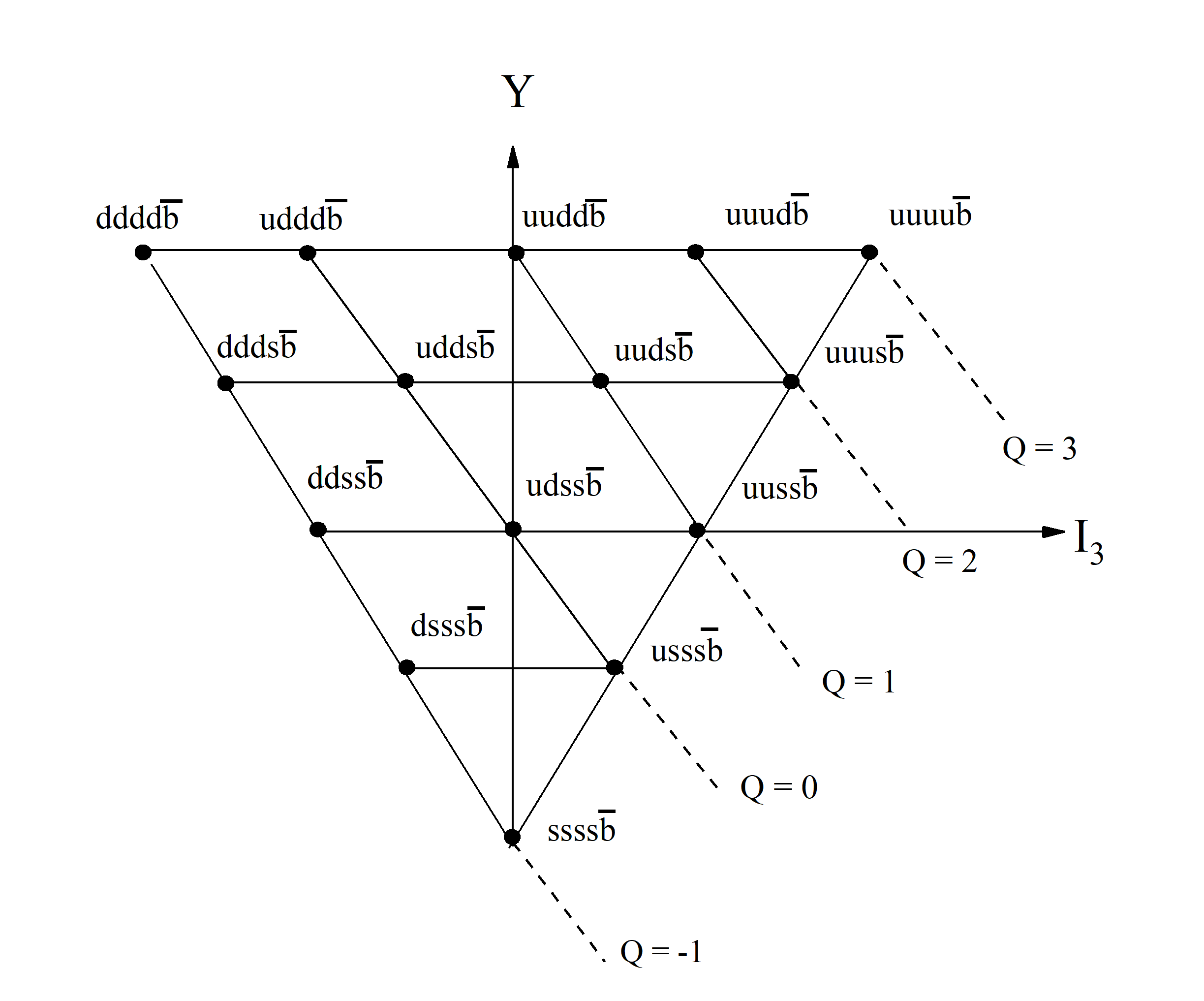}
\label{fig:2}
\end{figure}

In this subsection, we present a detailed analysis of the masses and magnetic moments of the 15-plet symmetric representation for pentaquarks with a single bottom quark, denoted by the quark content $qqqq\Bar{b}$. This study utilizes the extension of the Gursey-Radicati (GR) mass formula, the effective mass scheme, and the screened charge scheme to achieve a comprehensive understanding of these exotic hadronic states. A pictorial representation of the $Y-I_3$ plot for the 15-plet representation of pentaquarks with single bottom quarks are shown in Figure \ref{fig:2}. The masses of pentaquarks with single bottom quarks using the extension of the GR mass formula and the effective mass scheme are reported in Table \ref{tab:III}. Furthermore, the expressions for magnetic moments and corresponding assignments are written in Tables \ref{tab: expressions2} and \ref{tab: magnetic moments2}. The calculated masses and magnetic moments for the 15-plet symmetric pentaquark states with a single bottom quark reveal several key features. The masses exhibit a pattern consistent with the expected spin-spin interactions and symmetry considerations. The results offer predictions for yet unobserved states. The magnetic moments of the pentaquark states provide valuable information on their internal structure and spin configurations. The effective mass and screened charge schemes ensure that these values accurately reflect the influence of the heavy bottom quark and the light quark system. Thus, analyzing the symmetric 15-plet pentaquark states with a single bottom quark using the extended GR mass formula, effective mass scheme, and screened charge scheme offers a comprehensive understanding of their masses and magnetic moments. These methods collectively enhance the precision of theoretical predictions, contributing to the broader knowledge of exotic hadronic states and guiding future experimental searches.

\begin{table*}\renewcommand{\arraystretch}{0.8}
 \tabcolsep 0.5mm      
\centering
\caption{Table for the masses of 15-plet representation of pentaquarks with a single bottom quark using the extension of the Gursey-Radicati mass formula, effective mass scheme.}
\begin{tabular}{cccc}
\hline
\hline
 States & \hspace{0.3cm} G-R mass formula(MeV) & \hspace{0.3cm} Effective mass scheme & \hspace{0.3cm} Reference \\
  \hline
  \\
$P_{b0}^{++}$,  $P_{b0}^{+}$,  $P_{b0}^{0}$,  $P_{b0}^{-}$,  $P_{b0}^{--}$  & 7044.56 $\pm$ 38.70 & 6738.94 & \\
  \\
  \hline
  \\
  $P_{b1}^+$, $P_{b1}^0$, $P_{b1}^-$, $P_{b1}^{--}$  & 7144.19 $\pm$ 38.29 & 6876.16 & \\
 \\
  \hline
  \\
 $P_{b2}^{0}$,  $P_{b2}^{-}$,  $P_{b2}^{--}$ & 7243.83 $\pm$ 38.07 & 7017.71 & \\
  
  \\
  \hline
  \\
  $P_{b3}^{-}$, $P_{b3}^{--}$  & 7343.46 $\pm$ 38.04  & 7163.61 & \\
  \\
  \hline
  \\
  $P_{b4}^{--}$ & 7443.09 $\pm$ 38.21  & 7313.85 & \\
    \\
    \hline
    \hline
 \end{tabular}
   \label{tab:III}
\end{table*}

\begin{table}[ht]
    \centering
    \caption{Expressions for the magnetic moments of spin-parity equals to $5/2^-$ pentaquarks with a single bottom quark using effective quark masses (in $\mu_N$ ).}
    \begin{tabular}{cc}
    \hline
    \hline
\tabcolsep 0.5mm
 States  & \hspace{0.3cm} Magnetic moment expressions \\
 \hline
$P_{b0}^{++}$  & \hspace{0.3cm} 4 $\mu_u^{eff}$ + $\mu_{\Bar{b}}^{eff}$
\\
$P_{b0}^{+}$ & \hspace{0.3cm} 3$\mu_u^{eff}$ + $\mu_d^{eff}$ +  $\mu_{\Bar{b}}^{eff}$ \\
$P_{b0}^{0}$ & \hspace{0.3cm} 2$\mu_u^{eff}$ + 2$\mu_d^{eff}$ + $\mu_{\Bar{b}}^{eff}$ \\
$P_{b0}^{-}$ & \hspace{0.3cm} $\mu_u^{eff}$ + 3 $\mu_d^{eff}$ + $\mu_{\Bar{b}}^{eff}$  \\
$P_{b0}^{--}$ & \hspace{0.3cm} 4$\mu_d^{eff}$ + $\mu_{\Bar{b}}^{eff}$  \\
$P_{b1}^{+}$ & \hspace{0.3cm} 3$\mu_u^{eff}$ + $\mu_s^{eff}$ + $\mu_{\Bar{b}}^{eff}$ \\
$P_{b1}^{0}$ & \hspace{0.3cm} 2$\mu_u^{eff}$ + $\mu_d^{eff}$ + $\mu_s^{eff}$ + $\mu_{\Bar{b}}^{eff}$   \\
$P_{b1}^{-}$ & \hspace{0.3cm}  $\mu_u^{eff}$ + 2 $\mu_d^{eff}$ + $\mu_s^{eff}$ + $\mu_{\Bar{b}}^{eff}$  \\
$P_{b1}^{--}$ & \hspace{0.3cm} 3 $\mu_d^{eff}$ + $\mu_s^{eff}$ + $\mu_{\Bar{b}}^{eff}$  \\
$P_{b2}^{0}$ & \hspace{0.3cm}  
 2$\mu_u^{eff}$ + 2$\mu_s^{eff}$ + $\mu_{\Bar{b}}^{eff}$ \\
$P_{b2}^{-}$ & \hspace{0.3cm}  $\mu_u^{eff}$ + $\mu_d^{eff}$ + 2$\mu_s^{eff}$ + $\mu_{\Bar{b}}^{eff}$ \\
$P_{b2}^--$ & \hspace{0.3cm} 2 $\mu_d^{eff}$ + 2$\mu_s^{eff}$ + $\mu_c^{eff}$ + $\mu_{\Bar{b}}^{eff}$ \\
$P_{b3}^{-}$ & \hspace{0.3cm} $\mu_u^{eff}$ + 3$\mu_s^{eff}$ + $\mu_{\Bar{b}}^{eff}$ \\
$P_{b3}^{--}$ & \hspace{0.3cm} $\mu_d^{eff}$ +  3$\mu_s^{eff}$ + $\mu_{\Bar{b}}^{eff}$ \\
$P_{b4}^{--}$ & \hspace{0.3cm}  4$\mu_s^{eff}$ + $\mu_{\Bar{b}}^{eff}$ \\
\hline
\hline
\end{tabular}
  \label{tab: expressions2}
\end{table}

\begin{table*}[ht!]
    \centering
    \caption{Magnetic moments of pentaquarks with single bottom quark having $J^P = 5/2^-$ using the effective mass scheme, screened charge scheme, and effective mass and screened charge schemes together. All magnetic moment assignments are in the unit of $\mu_N$.}
    \begin{tabular}{|p{1cm}|c|c|c|c|}
    \hline
 States  &  Quark Content &  Effective mass Scheme  &  Screened Charge Scheme &  Effective mass + Screened Charge scheme\\
     \hline
$P_{b0}^{++}$ &  $uuuu\Bar{b}$ & 5.98 & 9.73 & 8.39\\
      
$P_{b0}^{+}$ &  $uuud\Bar{b}$ & 3.76 & 7.00 & 6.04 \\
       
$P_{b0}^{0}$ &  $uudd\Bar{b}$ & 1.54 & 4.28 & 3.69 \\
         
$P_{b0}^{-}$ &  $uddd\Bar{b}$ & -0.67 & 1.56 & 1.34\\
         
$P_{b0}^{--}$ &  $dddd\Bar{b}$ & -2.90 & -1.16 & -1.01 \\
          
$P_{b1}^{+}$ &  $uuus\Bar{b}$ & 4.03 & 7.21 & 6.33\\
           
$P_{b1}^{0}$ &  $uuds\Bar{b}$ & 1.77 & 4.21 & 3.68 \\
            
$P_{b1}^{-}$ &  $udds\Bar{b}$ & -0.47 & 1.21 & 1.04 \\
             
$P_{b1}^{--}$ & $ddds\Bar{b}$ & -2.73 & -1.78 & -1.59 \\             
$P_{b2}^{0}$ & $uuss\Bar{b}$ & 2.03 & 4.60 & 4.10 
\\
 $P_{b2}^{-}$ &  $udss\Bar{b}$ & -0.26 & 1.32 & 1.16
 \\
$P_{b2}^{--}$ &  $ddss\Bar{b}$ &  -2.55 &  -1.95 & -1.78 \\
            
$P_{b3}^{-}$ &  $usss\Bar{b}$ & -0.02 &  1.89 & 1.70 \\
             
$P_{b3}^{--}$ & $dsss\Bar{b}$ & -2.35 & -1.66  & -1.54 \\
              
$P_{b4}^{--}$ & $ssss\Bar{b}$ & -2.14 & -0.90 & -0.86\\
\hline
\end{tabular}
    \label{tab: magnetic moments2}
\end{table*}

\section{Production modes and decay channels of pentaquarks with a single heavy quark.}
One of the significant production mechanisms for pentaquarks containing a single heavy quark is through the decay of heavy baryons. The heavy baryon decays into a pentaquark state and a light meson. Heavy baryons contain at least one heavy quark, such as charm or bottom quark. These baryons can decay through weak interactions, which provide a pathway for producing exotic states such as pentaquarks. For example, heavy baryons such as $\Lambda_c$ ($udc$) or $\Lambda_b$ ($udb$) contain at least one heavy quark. These baryons can decay weakly, leading to the production of pentaquarks. The weak decay processes involve the transformation of one quark into another via the emission of a $W$ boson, followed by QCD-governed hadronization to form pentaquark states. This mechanism provides a rich avenue for studying exotic hadrons and understanding the dynamics of quark interactions in high-energy physics. Also, singly heavy pentaquarks can decay through strong interaction channels. These decay modes often involve the rearrangement of quarks into lighter hadrons. Thus, in the future, we can look into these production modes and decay channels for pentaquarks at experimental facilities like LHCb, etc. The detailed production mechanism and a few of the possible strong decay channels for 15-plet particles are shown in Table \ref{tab: Production Modes}.

\begin{table*}[ht]
    \centering
    \caption{The Table illustrates the production modes and strong decay channels of the 15-plet singly heavy pentaquarks. Here, $Q$ denotes a heavy quark, such as a charm or bottom quark, and $M(Q)$ represents the corresponding heavy meson. The production modes are considered in the context of the weak decay of heavy baryons. Additionally, the strong decay channels of the singly heavy pentaquarks are identified, where they decay into a light baryon and a heavy meson.}
    \tabcolsep 1.4cm
    \begin{tabular}{|c|c|c|}
    \hline
  Quark Contents &  Production Modes  &  Strong Decay Channels \\
     \hline
  $uuuu\Bar{Q}$ & $\Xi_{cc}^{++} \rightarrow uuuu\bar{Q} + \bar{D^0}$ & $uuuu\bar{Q} \rightarrow \Delta^{++} + M(Q)$ \\
      
 $uuud\Bar{Q}$ & $\Sigma_{c}^{++} \rightarrow uuud\bar{Q} + \bar{D^0}$ & $uuud\bar{Q} \rightarrow p + M(Q)$ \\
       
$uudd\Bar{Q}$ & $\Xi_{cc}^{0} \rightarrow uudd\bar{Q} + \bar{D^+}$  &  $uudd\bar{Q} \rightarrow n + M(Q)$ \\
         
 $uddd\Bar{Q}$ & $\Xi_{c}^{+} \rightarrow uddd\bar{Q} + \bar{D^+}$ & $uddd\bar{Q} \rightarrow \Sigma^- + M(Q)$ \\
         
$dddd\Bar{Q}$ & $\Omega_{c}^0 \rightarrow dddd\bar{Q} + \bar{D^0}$ & $dddd\bar{Q} \rightarrow \Delta^- + M(Q)$ \\
          
 $uuus\Bar{Q}$ & $\Xi_{cc}^{+} \rightarrow uuus\bar{Q} + \bar{D^+}$ &  $uuus\bar{Q} \rightarrow \Sigma^+ + M(Q)$ \\
           
$uuds\Bar{Q}$ & $\Xi_{c}^{+} \rightarrow uuds\bar{Q} + \bar{D^+}$ & $uuds\bar{Q} \rightarrow \Lambda + M(Q)$ \\
            
$udds\Bar{Q}$ & $\Xi_{c}^{0} \rightarrow udds\bar{Q} + \bar{D^+}$ & $udds\bar{Q} \rightarrow n + M(Q)$ \\
             
 $ddds\Bar{Q}$ & $\Omega_{c}^0 \rightarrow ddds\bar{Q} + \bar{D^0}$ & $ddds\bar{Q} \rightarrow \Sigma + M(Q)$ \\
              
$uuss\Bar{Q}$ & $\Omega_{c}^0 \rightarrow uuss\bar{Q} + \bar{D^0}$ & $uuss\bar{Q} \rightarrow \Sigma^+ + M(Q)$ \\

$udss\Bar{Q}$ & $\Xi_{c}^{0} \rightarrow udss\bar{Q} + \bar{D^+}$ & $udss\bar{Q} \rightarrow \Sigma^0 + M(Q)$ \\
           
$ddss\Bar{Q}$ & $\Xi_{c}^{0} \rightarrow ddss\bar{Q} + \bar{D^+}$ & $ddss\bar{Q} \rightarrow \Sigma^- + M(Q)$ \\
            
$usss\Bar{Q}$ & $\Xi_{c}^{+} \rightarrow usss\bar{Q} + \bar{D^+}$  & $usss\bar{Q} \rightarrow \Omega^- + M(Q)$\\
             
$dsss\Bar{Q}$ & $\Xi_{c}^{0} \rightarrow dsss\bar{Q} + \bar{D^0}$ & $dsss\bar{Q} \rightarrow \Xi^- + M(Q)$ \\
              
$ssss\Bar{Q}$ & $\Omega_{c}^0 \rightarrow ssss\bar{Q} + \bar{D^0}$ & $ssss\bar{Q} \rightarrow \Omega^- + M(Q)$ \\
\hline
\end{tabular}
    \label{tab: Production Modes}
\end{table*}

\section{Summary}
The recent experimental discoveries of singly heavy tetraquark structures have reignited interest in studying exotic hadronic states beyond the conventional quark model. These findings, particularly those from the LHCb collaboration, provide compelling evidence for the existence of multiquark states and challenge our understanding of Quantum Chromodynamics (QCD) in the non-perturbative regime. In light of these developments, exploring the properties of pentaquark states, especially those containing a single heavy quark (charm or bottom), has become a focal point of theoretical high-energy physics. Our investigation began with classifying pentaquark states according to their quark content and the corresponding SU(3) flavor symmetries. The symmetric 15-plet representation, which emerges from the fully symmetric arrangement of four light quarks and one heavy antiquark, served as the cornerstone of our analysis. The Gursey-Radicati mass formula is an essential tool for estimating the masses of hadrons based on their quark content and symmetry properties. The formula is extended to accommodate the 15-plet symmetric representation of pentaquarks, incorporating both light and heavy quarks. The extended GR mass formula considers contributions from the spin-spin interactions, quark masses, and other symmetry-breaking effects. To account for the dynamic interactions among quarks within the pentaquark system, we employ an effective mass scheme. The effective mass of each quark is modified by the surrounding quarks and the gluonic field, leading to a more accurate representation of the internal structure. The effective mass is calculated by incorporating the spin-spin interactions among the constituent quarks. Moreover, The screened charge scheme is used to refine the potential interactions between quarks. In a multiquark system, the effective charge of each quark is screened by the presence of other quarks, reducing the effective interaction strength. The magnetic moments were analyzed using three distinct schemes: the effective mass scheme, the screened charge scheme, and a combined approach incorporating both effects. Our findings revealed significant variations across these schemes, underscoring the complex interplay between quark masses and charge screening effects in determining the magnetic properties of pentaquarks. The combined scheme, which balanced the contributions from effective mass and screened charge corrections, provided the most realistic predictions, offering a valuable reference for future experimental investigations. In addition to calculating the masses and magnetic moments of pentaquarks containing a single heavy quark, we have also proposed potential production modes originating from the decay of heavy baryons. Furthermore, we have identified strong decay channels where these pentaquarks transition into a light baryon and a heavy meson. These findings not only deepen our understanding of the structure and dynamics of singly heavy pentaquarks but also pave the way for future experimental validations. The suggested production and decay mechanisms are crucial steps towards uncovering new exotic hadronic states and could significantly impact ongoing research in hadron physics. This comprehensive study thus provides a solid foundation for both theoretical advancements and experimental pursuits in the exploration of these novel states. In conclusion, our study of singly heavy pentaquarks has not only expanded the understanding of these exotic states but has also demonstrated the efficacy of various theoretical models in predicting their properties. The insights gained from this work will serve as a foundation for future research in the field of hadron spectroscopy, particularly in exploring pentaquarks with heavy quarks. By analyzing the properties of $qqqq\bar{Q}$ pentaquarks, such as their masses, magnetic moments, and decay channels, one can gain deeper insights into the strong force and the confinement mechanism that binds quarks together. As experimental techniques continue to advance, the predictions made in this study will be crucial in guiding the search for these fascinating particles, contributing to the broader effort of uncovering the full spectrum of hadronic states within the framework of Quantum Chromodynamics (QCD).

\nocite{*}

\bibliography{15}

\end{document}